# Explainable OOHRI: Communicating Robot Capabilities and Limitations as Augmented Reality Affordances

**Lauren W. Wang**
Princeton University
Princeton, USA
wlauren@princeton.edu

**Mohamed Kari**
Princeton University
Princeton, USA
mokari@princeton.edu

**Parastoo Abtahi**
Princeton University
Princeton, USA
parastoo@princeton.edu

## Abstract

Human interaction is essential for issuing personalized instructions and assisting robots when failure is likely. However, robots remain largely black boxes, offering users little insight into their evolving capabilities and limitations. To address this gap, we present explainable object-oriented HRI (X-OOHRI), an augmented reality (AR) interface that conveys robot action possibilities and constraints through visual signifiers, radial menus, color coding, and explanation tags. Our system encodes object properties and robot limits into object-oriented structures using a vision-language model, allowing explanation generation on the fly and direct manipulation of virtual twins spatially aligned within a simulated environment. We integrate the end-to-end pipeline with a physical robot and showcase diverse use cases ranging from low-level pick-and-place to high-level instructions. Finally, we evaluate X-OOHRI through a user study and find that participants effectively issue object-oriented commands, develop accurate mental models of robot limitations, and engage in mixed-initiative resolution.

## CCS Concepts

• **Human-centered computing** → **Mixed / augmented reality**;
• **Computer systems organization** → **Robotics**.

## Keywords

Human-robot interaction (HRI), Augmented reality (AR)

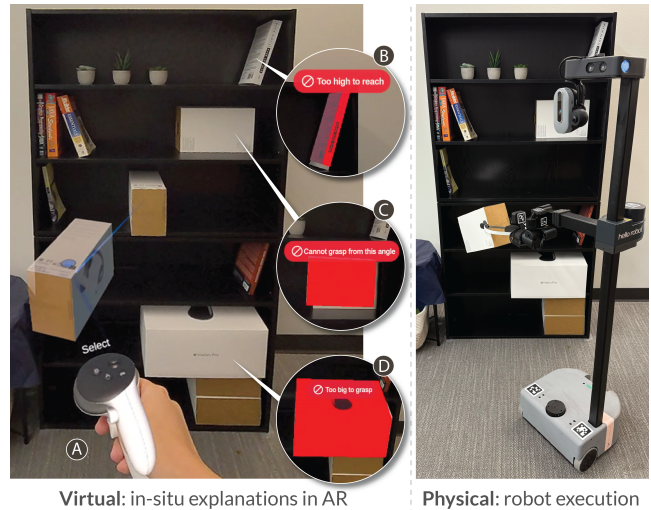

**Virtual**: in-situ explanations in AR   **Physical**: robot execution

Figure 1: X-OOHRI conveys action possibilities and constraints through an AR interface. The user reorganizes a shelf by directly manipulating the virtual twin of a box to specify placement (A). When the user attempts to move other objects, the system blocks the action for a book on top due to a height constraint (B), a box in an ungraspable orientation (C), and a larger box due to its size (D). The system visualizes these constraints using color coding and explanation tags. After confirmation, a robot manipulator executes the instructions.

## 1 Introduction

We are increasingly surrounded by autonomous robots that augment our capabilities in applications such as house cleaning, package delivery, or security monitoring [85]. Such applications require users with limited technical expertise to specify personalized instructions and help robots when they inevitably fail under complex real-world conditions [11]. However, during these interactions, users often struggle to determine what the robot can or cannot do [23, 52]. As a result, they cannot take advantage of the robot's full capabilities and do not understand why some actions fail. Moreover, for robots that learn through real-world interactions, capabilities and limitations change over time. Even after establishing common ground [49], users are challenged due to outdated mental models.

Existing work has focused on generating textual explanations for failures [14, 23, 26]. However, language-based interfaces are prone to ambiguity and discoverability issues [21]. Augmented reality (AR) has emerged as a promising approach for human-robot interaction [76], including end-user robot programming [16, 27, 28, 40, 54, 66] and communicating robot intent [69, 77, 86]. Yet most AR systems are used either solely for giving instructions [15, 17, 24, 40] or visualizing robot states [36, 69, 77]. We identify a gap in using AR to provide in-situ explanations during direct instructions that take the task context and world state into account.

We introduce explainable object-oriented HRI (X-OOHRI), which uses AR to communicate robot capabilities and limitations in situ. It enables non-technical users to explore action possibilities by directly interacting with virtual twins of real objects colocated in the physical environment and receiving real-time feedback when actions are infeasible.

To explain the bounds of what the robot can or cannot do, we present five communication channels. *Signifiers* help users identify which objects are actionable. A *radial menu* presents the action possibilities for selected objects. Real-time red *color coding* indicates a limitation, accompanied by an *explanation tag* to communicate the constraint. Users can visualize system-driven resolutions as *preview animations* in AR to understand available recovery strategies.

We draw inspiration from object-oriented programming [9, 48]. Physical objects and robots are represented as class instances with defined attributes and methods generated by a VLM. These OO abstractions are used in a simulated environment to infer action feasibility via a decision tree and create constraint explanations. After resolving constraints, commands are sent to the robots for execution.

To demonstrate the usability of X-OOHRI, we present scenarios with physical robots that showcase the full pipeline from instruction and explanation visualization to robot execution. Finally, we conducted a study in which participants completed task sequences, showed understanding of robot capabilities and limitations, and rated the system as helpful and easy to use.

**Contributions:**
- A novel HRI paradigm leveraging AR visualizations and communication channels for generating explanations and facilitating mixed-initiative resolutions.
- A full implementation of the X-OOHRI system, including VLM-generated object-oriented classes, multi-robot affordance matching, and physical robot integration.
- Technical demonstrations[1] and a user study as evaluations.

## 2 Related Work
### 2.1 Explainable Robotics
Research has shown that humans and robots often lack shared mental models and situational awareness, leading to misunderstandings and low agreement [62], thereby giving rise to the field of explainable robotics [4, 35, 73]. Explanations differ in quantity, quality, and modality. Real-time displays are critical for explanations [32]. Kim et al. [50] emphasized that users prioritize actionable insights over technical system details. Rosentahl et al. [70] and Wang et al. [81] demonstrated the use of natural language for explanations.

Error explanations enable the surfacing, recovery, and prevention of failures [3, 23]. To surface failure, Duan et al. [26] proposed using a VLM for detection. To recover from failure, Das et al. [23] proposed auto-generating explanations amidst execution. However, to prevent failures in the first place, a robot needs to communicate its situation-specific competencies [14, 39, 41, 68, 75]. Such failure-preventive explanations require self-assessment that determines the robot's limits and potential failure modes [18, 19, 30, 33, 71]. Saycan [1] introduced environmental context to an LLM pretrained on robot skills. Israelsen et al. [41] and Gautam et al. [30] emphasized alignment with the environment and the robot's limits as measures of competency.

As explanations of robot capabilities and limitations lead to more appropriate user reliance [20], we also aim to provide such explanations. However, in contrast to previous work, following our object-oriented paradigm, we map robot capabilities and limitations directly to object affordances. This enables object-oriented self-assessment ranging from single-object operations to dynamic sequences of multi-object manipulations.

### 2.2 Augmented Reality for HRI
Augmented reality has emerged as an effective medium for communicating a robot's state to the user. In particular, AR has proven useful to present the robot's motion intent [63, 76] by displaying trajectories [69, 77, 86]. AR has also been used to highlight physical entities relevant to the robot action [34]. Furthermore, AR has communicated social cues from the robot to the user by overlaying characters [25] and to support robots' local affordance reasoning [17, 65].

User-to-robot communication predominantly aims at robot programming where the user issues commands to the robots. This process requires addressing challenges in interacting with the environment and task-relevant objects [2]. Early approaches used projections on physical objects [5, 10, 72]. Later work used wearable or handheld AR to improve teleoperation through trajectory modification [8, 24, 28, 36, 66]. Systems such as RoboVisAR [54] and V.Ra [16] simplified motion path creation and task planning using AR. GhostAR [15] introduced demonstrative role-playing and spatiotemporal editing of human ghosts to program robots. While these works highlight AR's potential for instructions, establishing truly bidirectional communication requires spatial comprehension of physical objects beyond proxies and panels [12, 37, 38, 80].

"Digital twins" and "virtual twins" have emerged in recent years. Research shows that integrating situated visualizations on virtual counterparts with their physical referents is effective for tasks involving physical interaction [51, 53, 58–60, 82]. Marcer [40] used natural language as the instruction input and AR to display virtual outcomes. X-HRTraining [78] demonstrated programming by demonstration or previews with virtual objects. HoloSpot [29] used a World-in-Miniature (WiM) interface that allowed users to drag and drop a virtual object to issue pick-and-place commands.

In contrast to prior work, X-OOHRI supports direct manipulation of multiple colocated, life-size, and interactive virtual twins. More distinctly, instead of using AR solely for interactivity, we leverage it to visualize explanations for robot tasks. By encoding object-oriented affordances into virtual twins, the system generates explanations of robot capabilities grounded in object properties.

### 2.3 Object-Oriented Programming
Object-oriented programming (OOP) is a widely recognized paradigm in programming languages that structures data into classes, encapsulating attributes as fields and behaviors as methods, distinguished by key features such as data abstraction, message passing, modular programming, polymorphism, and inheritance [9, 44, 48].

OOP has been adopted in robotics research for its structured approach to data encapsulation, enabling rapid software prototyping for robots [13]. Additionally, it has been applied to assembly tasks by modeling parts as objects with unique properties and implementing operational procedures as class methods [6, 7]. OOP also supports modular design by treating robot components, such as end-effectors, as objects [31]. Beyond physical systems, it has been employed in simulations [67] and human-like learning processes by structuring object-oriented knowledge from robot experiences [45].

---
[1] Video demonstrations available on: https://xoohri.github.io/

Instead of using the object-oriented paradigm as a tool for software optimization or performance gains, we introduce it as a novel paradigm for human-robot interaction. Our work differs from prior extensions of the OO concept in robotics by applying it to an AR interface with robot-mediated physical actions, thereby shifting the locus of interaction from robots to objects. In our approach, object-orientedness serves a dual purpose. First, it supports end-user robot instruction through direct manipulation and task assignment to objects. Second, it encodes VLM-generated object affordances and robot capabilities in an OO schema, enabling on-the-fly inferences about action possibilities and explanation generation.

## 3 X-OOHRI

X-OOHRI enables users to instruct robots by directly interacting with virtual twins, using AR communication channels to present capabilities and limitations, visualize explanations derived from encoded object affordances, and offer resolutions when issues arise.

### 3.1 *GhostObject* Interactions

Humans derive action possibilities from their perception of objects in the environment [22, 61]. For example, a pile of boxes can be moved, and a floor can be vacuumed. Since these affordances define how robots can act on objects, we envision an interaction paradigm that enables users to give robot instructions by acting directly on spatially situated objects in AR. To achieve this, we implement our previously introduced *GhostObjects* [79]—life-size, world-aligned virtual twins of real-world objects. Users can select targets via raycasting or lasso selection, then manipulate *GhostObjects* by moving, scaling, aligning, or altering them to meet task requirements. Placement can be specified by directly dragging the object, with a snap-to-default feature for quick repositioning. Beyond interaction, *GhostObjects* can enable visual representations to dynamically convey robot action capabilities, limitations, and potential resolutions at both the individual and multi-object levels.

### 3.2 Communication Channels

Effective human–robot interaction requires understanding which actions are available, what constraints may arise, and how to resolve them in situ, which X-OOHRI communicates through AR cues.

A white circle **Signifier** appears when the controller's ray intersects an object with encoded action possibilities, making interaction opportunities explicit. After selecting an object, users can perform the default *Move* action by repositioning its virtual twin to set the placement spatial parameter, or choose other VLM-generated actions (e.g., *Stack*, *Fetch*, *Dust*) from a **radial action menu**, with composite actions decomposed into sub-steps through preconfiguration. Limitations are communicated on two levels: actions that are infeasible given the object, robots, or current world state appear grayed out, and actions absent from the menu implicitly indicate that the object has no such affordance. For example, *Stack* will not appear for spherical objects, and *Fetch* will appear grayed out for graspable objects that are obstructed, indicating a potentially resolvable limitation. **Color coding** denotes binary object states dynamically applied to virtual twins: red indicates an action limitation and reverts to its original color once the constraint is resolved. For example, if an object cannot be picked by any of the available robots, its *GhostObject* appears red and immovable; if picking is possible but the placement is invalid, the red highlight updates dynamically as the user moves it. **Explanation tags** complement color coding by labeling constraint causes and can expand into detailed tooltips, with content initialized from common errors and extended as new failure types arise. Choosing a resolution may trigger a **Preview animation** that lerps through an action sequence, with complex cases authored manually. Together, these cues make capabilities discoverable, limitations visible, and resolutions actionable in situ.

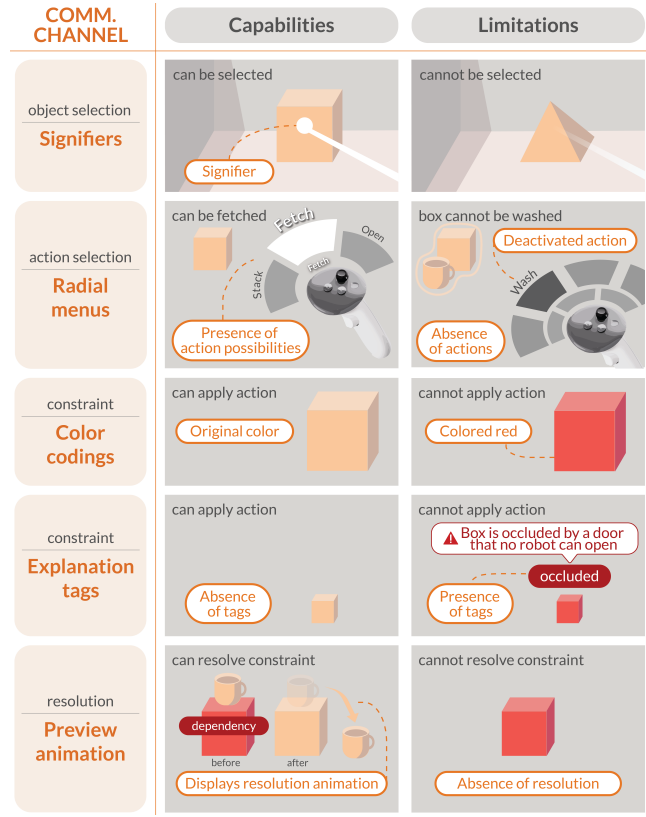

Figure 2: Channels for communicating robot capabilities and limitations: signifiers, radial action menus, *GhostObject* color coding, explanation tags, and animated resolution previews.

### 3.3 Object-Oriented Schema

X-OOHRI extends the OO paradigm beyond spatial interaction to its backend, encoding capabilities and limitations within object and robot classes. This structure jointly evaluates the attributes of both classes to determine action feasibility. We describe these two classes below. **XObject class** represents real-world items, with attributes such as size, weight, position, and surface properties collectively determining available actions. Intrinsic attributes (e.g., size, weight) are immutable, while dynamic states (e.g., position, rotation) update at runtime. Methods define the object's action possibilities (e.g., all mugs may be moved, filled, or stacked) while attribute values can enable or restrict specific actions. Classes with similar capabilities inherit from shared base classes, supporting multi-level abstraction. Users can also define and store action sequences as custom methods. **XRobot class** represents robots with static capabilities (e.g., payload, movement range) derived from technical specifications and dynamic properties (e.g., position, battery life). Methods define available actions (e.g., pick, place) and operate polymorphically across compatible objects within the robot's limits. We generate XObject instances by passing images to a VLM with an abstract class template, and similarly generate XRobot instances from technical datasheets. This structure also supports software updates, allowing newly acquired robot capabilities to be incorporated as they become available. To ensure controlled study conditions, we manually verified the VLM outputs for our experimental setup. For deployments outside the lab, automated verification using execution feedback or language-model-based self-validation [56, 83] may be useful but is left for future work.

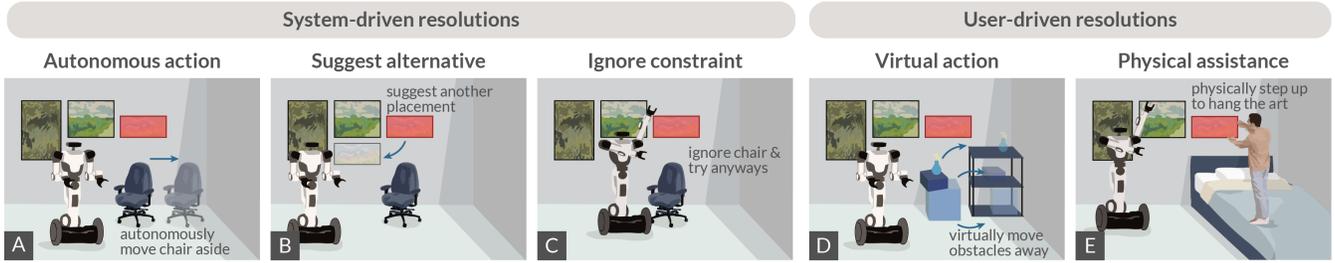

Figure 3: Five resolution strategies illustrated in a scenario where the user specifies hanging a small poster on the right side of the wall, but the task is infeasible due to obstructions such as a chair, stacked items, or a bed.

### 3.4 Formulating Affordances and Explanations

With object and robot properties encoded, the system can detect when a task instruction is infeasible and display relevant explanations. Feasibility is implemented in the action module (Figure 5), determined through a decision tree defined as $f : R_A \times O_A \times S \rightarrow e$, where $R_A$ is the set of action capabilities of available robots, $O_A$ is the set of action possibilities for the selected object, and $S$ denotes the world state, represented in a Unity-simulated environment with a reach mesh derived from robot limits and scene data. The output $e$ denotes the failure condition. For example, in manipulative tasks, the decision tree evaluates reachability, graspability, and manipulability. To determine reachability, the module locates the object in the simulated environment and checks for potential occlusions, and whether the robot can navigate to it and reach it given its joint limits. Graspability is assessed by comparing the object's size and orientation attributes with the gripper's aperture and global pose, while manipulability can involve verifying that the object's weight does not exceed the gripper's payload limit. When infeasibility is detected, the system raises an explanation $\varepsilon(e)$ that identifies the limiting preconditions and updates the tag label. For example, displaying a "Too high to reach" explanation tag when the height condition fails. For each selected object, the action module checks through available robots to evaluate the feasibility of specified actions based on varying criteria. For example, a vacuum action does not require grasping but must verify navigation based on the robot's capabilities, the selected floor area, and the surrounding environment.

### 3.5 Mixed-Initiative Resolution Strategies

When limitations arise, X-OOHRI supports system-driven and user-driven resolutions. System-driven resolutions appear on the radial menu as Auto, Alternative, and Ignore. Motivated by autonomous strategies from prior robotics research [18, 19, 30, 33, 71], the **Auto** resolution addresses constraints by manipulating other objects in the environment to make the specified action feasible, while leaving the target object and its action unchanged. When selected, the system plays a preview of the action sequence. Inspired by prior work on reach-avoid scenarios [42], the **Alternative** resolution modifies the instruction by suggesting either a different object of the same class or another target pose. When selected, the alternative object flashes briefly, or the action is previewed at the new pose. Auto and Alternative appear grayed out when no corresponding resolution exists. In cases where the proposed resolutions are undesirable, or the user does not consider the constraint problematic, they may choose to **Ignore** the constraint and let the robot attempt execution. Users can also resolve constraints themselves. They may issue additional **Virtual actions** to add preparatory steps or provide **Physical assistance** outside AR when the robot is incapable of resolving the issue.

## 4 Implementation

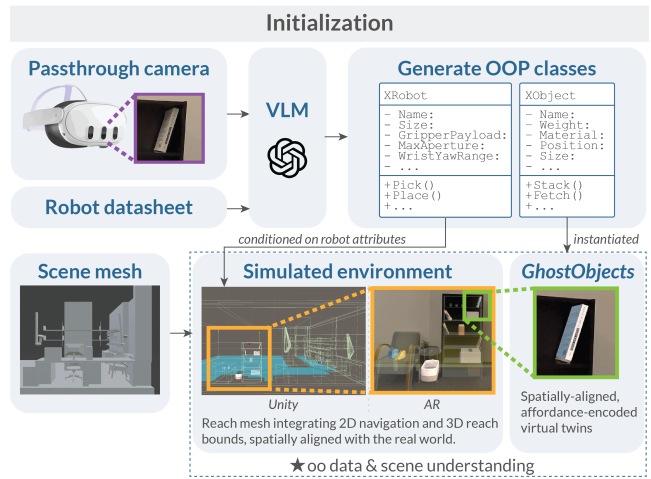

Figure 4: System initialization: scene capture and robot specification sent to a VLM to generate OO affordances, instantiate *GhostObjects*, and construct a simulated environment spatially aligned with the real world.

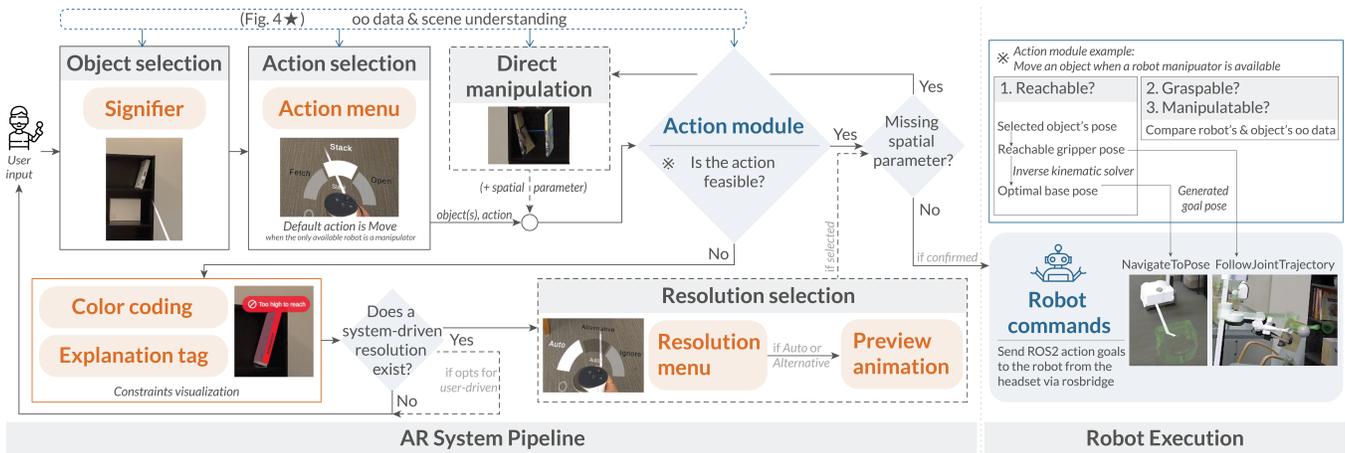

Figure 5: The AR pipeline begins with object and action selection via raycasting and the radial menu. The action module validates feasibility using user input, object-oriented data, environment information, and optional spatial parameters. If limitations arise, the system updates color coding, explanation tags, and optional resolution strategies for users to preview and confirm, before publishing ROS2 messages with generated poses for robot execution.

## 4.1 System Setup and Initialization

Our AR system was deployed as a standalone application on a passthrough head-mounted display (Meta Quest 3), chosen for its wider field of view compared to AR glasses and easier 3D depth manipulation compared to 2D touchscreen devices. The implementation used Unity (2022.3.56f1), Blender (v4.3), and the Meta XR All-in-One SDK (v74). AR interactions were designed for one-handed use with the Quest controllers, relying only on the trigger button and thumbstick; the left-hand controller functioned as a textual panel with developer toggles. Images captured from the headset's cameras via the Meta passthrough API were processed by a VLM (GPT-5.1) to generate object-oriented structures in advance, avoiding latency during real-time interaction. Environment reconstruction employed the headset's Space Setup function to scan the scene and generate bounding boxes of spatial anchors. Anchor positions from Meta's Mixed Reality Utility Kit (MRUK) were used as reference points for spawning virtual objects and bounds, ensuring persistent alignment without extensive recalibration. High-precision virtual meshes were modeled manually in Blender to match the physical scene for accurate object manipulation. Publicly available image-to-3D models and 3D scans from the Stretch 3 robot's depth camera were tested but did not meet our requirements; however, future reconstruction methods could be substituted as needed. The *GhostObjects* [79] were then instantiated with encoded VLM-generated OO affordances. Using high-fidelity scene meshes and robot's physical reach limits derived from its attributes, we generated a 2D navigation terrain with Unity NavMesh and combined it with the 3D scene meshes to create reachability bounds. Together, these constructed a simulated environment with virtual objects, spatially aligned with the real world.

## 4.2 AR System Pipeline

The interaction begins when the user raycasts into the environment with the controller. Signifiers indicate objects with encoded action possibilities. The user can then view and choose from the actions on a radial menu. We made the default action *Move*, as pick-and-place underlies most modern robot capabilities. In a *Move* example, when the user attempts to drag an object, the system queries the Action module to check available robots with moving capability. In our demonstration, a mobile manipulator robot is available, so the module proceeds to determine if *Pick()* is possible by checking whether the object is (1) reachable, (2) graspable, and (3) manipulatable. Using object, robot, and environment data, the module verifies that the object is within the reach mesh, fits within the gripper's aperture, weighs less than the maximum payload, etc. If all conditions are satisfied, the user can freely reposition the object. Because the *Move* action requires a spatial parameter—placement—the module uses the target pose specified through direct manipulation to evaluate the *Place()* condition for reachability. If a condition fails, the system applies red color coding to indicate a binary 'feasible/not feasible' state and attaches an explanation tag with a short label drawn from a global list informed by the failure condition. These constraint visualizations update dynamically as the user's input changes. Users can resolve them through mixed-initiative strategies. System-driven options appear in the radial menu, where Auto and Alternative trigger a preview animation, and selecting one unlocks the action to proceed. Users may also issue a different instruction or provide physical assistance.

## 4.3 Robot Execution

After constraints are resolved and the user confirms, the system sends ROS2 action goals to the robot. For demonstration, we implemented X-OOHRI on a Hello Robot Stretch 3, controlled by a state machine built on Skynet [47] and communicated via ROSbridge. We run SLAM using the robot's LiDAR to build a 2D navigation map and use the on-robot AMCL localization package with a custom coordinate transformation method to align the robot and Unity headset frames. This enables the robot to navigate to the correct locations as viewed and specified in the headset without additional hardware.

Each object has predefined gripper poses. During feasibility checking, the Action module determines whether a reachable pose exists and uses inverse kinematics to compute the joint angles and arm configuration that yield the ideal base pose. The resulting pick and place poses are stored, and when a user issues a robot command, the system uses these poses to publish *NavigateToPose* and *FollowJointTrajectory* messages to the robot. If execution fails, the system logs the error, updates the red color coding, and attaches the error message as an explanation for the user to review later.

### 4.4 Technical Demonstration

We evaluated OO affordance generation with GPT-5.1 on 20 objects, achieving 98.7% categorical accuracy with minor numerical deviations and logically accurate methods. Robot execution reached an 80% success rate over 20 AR-instructed pick-and-place trials, with pose error (M=4.5 cm, SD=3.09 cm). AR spatial alignment error (M=7.48 cm, SD=1.58 cm) was corrected with brief recalibration (M=13.15 s, SD=2.83 s). End-to-end latency was measured (M=0.23 s, SD=0.15 s) (Appendix A). We demonstrated three scenarios covering single-object pick-and-place, multi-object manipulation, and composite actions. See video demonstrations.[1]

(1) **Book repositioning.** The user wants to place the book on a different shelf. She moves the virtual twin across locations to specify placement while understanding the height constraint. A robot then executes the pick-and-place action (Figure 1).

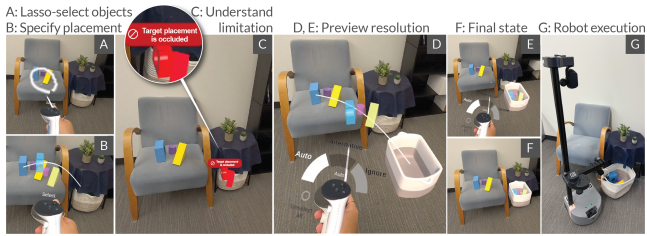

Figure 6: The user tries to store foam blocks into an occluded basket, sees the limitation, and opts for Auto resolution: moving the basket.

(2) **Room organization.** The user wants to tidy up foam blocks on an armchair. She lasso-selects the blocks and drags them into a nearby basket. Color coding and an explanation tag indicate that the placement is infeasible because the basket is occluded. She selects the Auto resolution, previewing an animation that moves the basket out, places the blocks inside, and returns it. After confirmation, a robot manipulator executes the action sequence (Figure 6).

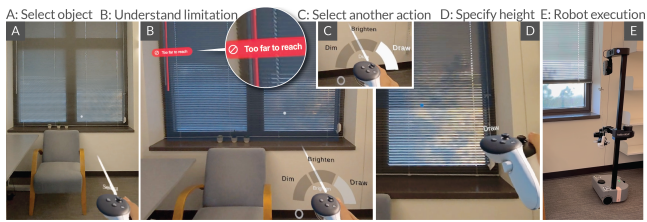

Figure 7: The user selects *Brighten*, understands its limitation, and instead chooses *Draw* to open the blinds by moving the controller up.

(3) **Blinds adjustment.** The user finds the room too dark and selects the blinds. He opens the radial menu and sees the *Brighten* action deactivated, with the tilt wand color-coded red and an explanation tag indicating it is unreachable. He instead chooses the *Draw* action to open the blinds by raising his hand to specify the height. After confirming, a robot manipulator executes the task (Figure 7).

## 5 User Evaluation

We conducted an IRB-approved preliminary study aiming to (1) assess users' understanding of the robot's capabilities and limitations, and (2) gather quantitative and qualitative feedback. As no comparable systems exist for visually communicating robot limitations via object-oriented interactions, we opted for an exploratory study rather than a controlled experiment. Fourteen participants (12 male, 2 female; aged 18–25) were recruited. Three had no AR experience, ten had limited experience, and one used AR weekly. In robotics, seven had no experience, six had some academic exposure, and one was a medical robotics expert. No one had used our system before.

### 5.1 Procedure

Participants were instructed to think aloud during the study while wearing an AR headset. The 40-minute session comprised two parts, each preceded by a tutorial on basic AR controls and communication channels. As the study focused on user understanding through OO instructions, all interactions occurred before execution; no physical robots were used, and virtual robots were shown only at the end of each task for confirmation. After completing both parts, participants filled out a five-part survey assessing self-reported and factual understanding of the robot's capabilities, limitations, and need for assistance, as well as task load (NASA-TLX), usability (SUS), and feedback on AR visualizations. All responses used a 5-point Likert scale (Appendix B). Finally, participants completed a 15-minute audio-recorded semi-structured interview and received a $16 gift card.

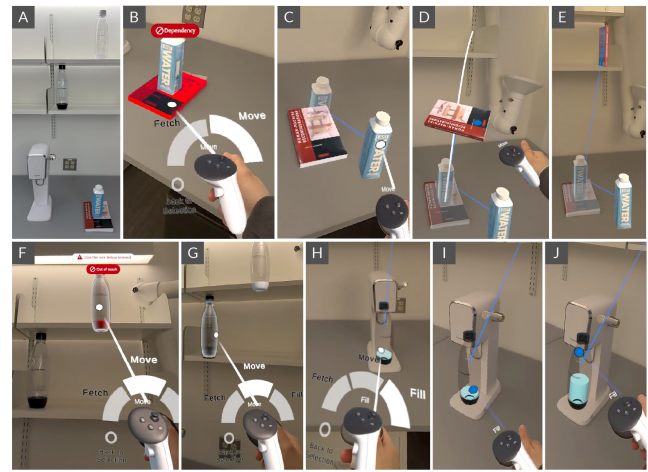

Figure 8: Part 1 task sequence: (A-E) moving an obstructed book by first moving the water carton aside to resolve dependency; (F-G) moving a bottle to the soda machine, accepting the system's alternative, and (H-J) specifying the fill amount.

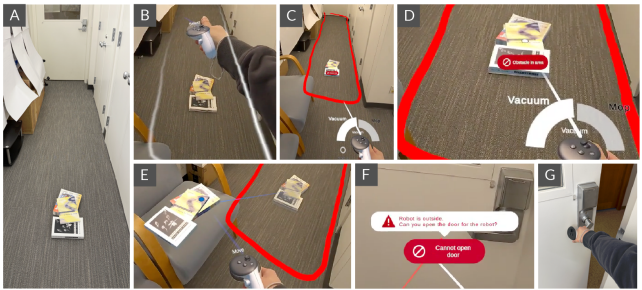

Figure 9: Part 2 task sequence: (A-D) specifying the vacuum area, (E) clearing obstacles, and (F-G) helping the vacuum robot enter the room by physically opening the door.

**Part 1.** Participants instructed a pick-and-place action and a fill action. Tasks included: (1) moving a book blocked by a water carton, resolved through a user-driven resolution by virtually repositioning the carton; (2) placing a bottle under the soda machine, where the top-shelf bottle was out of reach, prompting an explanation tag and selection of a system-driven alternative resolution using a lower-shelf bottle; and (3) filling it by directly adjusting the liquid level.

**Part 2.** Participants instructed a vacuum action on a desired area of the aisle, scattered with books. They (1) lasso-selected the target area and triggered vacuuming, (2) resolved an obstacle constraint by virtually moving the books out of the cleaning zone, and (3) addressed a navigation limitation by physically opening the door for the robot to enter. The virtual mobile manipulator then cleared the obstacles, and the virtual vacuum cleaned the floor.

## 5.2 Results

All participants successfully completed the tasks, despite some having limited AR experience, with AR interaction time per subtask (M=114 s, SD=53 s). Findings are presented, with numbers in parentheses indicating participant ID or count.

**Capabilities, limitations, and constraint resolutions.** All participants agreed the interactions helped them understand the robots' capabilities (6 strongly agree, 8 slightly agree), limitations (9 strongly agree, 5 slightly agree), and when to assist (11 strongly agree, 2 slightly agree, 1 neutral). Regarding capabilities, participants recalled object manipulation (14), vacuuming (14), and operating the soda machine (10). Several participants (7) named unused actions from the radial menu, indicating they recognized them as a display of the robots' capabilities. Additionally, participants inferred higher-level abilities such as perception, reasoning, object recognition, and self-assessment. As P14 noted, "*It could detect if objects were in collision. It could also map object locations.*" Limitations were recognized through communication channels, including the inability to vacuum under objects (9), open a closed door (7), manipulate items from the bottom of a stack (5), or reach beyond a range (5). All participants were able to resolve the constraints either proactively or after interacting with the explanation tags. They recalled virtually repositioning objects (12), physically opening doors (10), or selecting system-suggested alternatives (2). P13 commented, "*If there was a physical limitation, it would tell me what was needed.*"

**AR visualization and design suggestions.** Most participants found the AR visualizations clear (13) and accurate (12), highlighting features such as direct manipulation of *GhostObjects*, color coding, explanation tags, and the radial action menu. As P11 remarked, "*The visualization was amazing...pretty spot on.*" and P3 noted it was "*well integrated with real life.*" Suggestions included improving *GhostObject* visibility with distinct highlights, outlines, or animations (P11, P13), and graying out moved real objects (P6). Additional recommendations were to extend explanation tags beyond text (P4) and support free-hand interaction (P6).

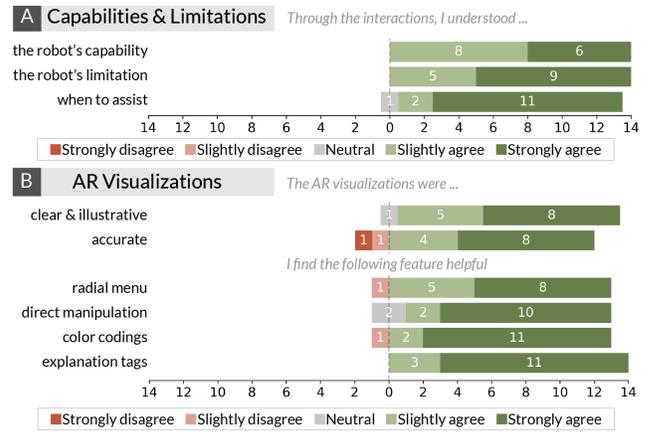

Figure 10: Self-reported participant ratings: (A) understanding robot capabilities/limitations and when to help, and (B) clarity, accuracy, and helpfulness of AR interactions and visualizations, including the communication channels.

**Task load and usability.** NASA-TLX results indicated low workload in all six indices, with slight increases in mental demand reported by participants due to AR unfamiliarity. The overall SUS score was 79.3 ("good" usability). Participants generally found the system easy to use, and many reported they would use it frequently, particularly for complex tasks. Some noted the inefficiency for short-distance movements that could be done faster manually (P10), while others highlighted its potential for users with limited mobility (P12).

## 5.3 Discussion

In our study, participants reported low task load, high usability, and a clear understanding of capabilities and limitations. We next discuss AR's role, action abstraction, and user control.

**Balancing AR advantages and trade-offs.** Augmented reality enhances users' expressive power and increases information bandwidth between humans and robots. However, unfamiliar technology like wearable AR introduces a learning cost: users must adapt to a new interface and interaction paradigm. In our study, 4 of 14 participants reported needing additional technical support to use the system effectively, indicating a need for improved onboarding. Additional guidance or more constrained interactions could mitigate these challenges, but may steer users toward predefined actions and trade off expressivity for support.

While recent vision–language–action foundation models have fueled enthusiasm for direct voice-to-action control, language interfaces alone suffer from discoverability and ambiguity issues. Our approach addresses these gaps by surfacing the *in-between* decision points—capabilities, limitations, and constraints—in a physically colocated, visually rich format. Rather than replacing other modalities, X-OOHRI can complement them, integrating with conditional triggers [40, 54], timeline-based task editing [15], or multimodal input [40] that uses language when it is efficient and AR when clarity, discoverability, and spatial reasoning are needed.

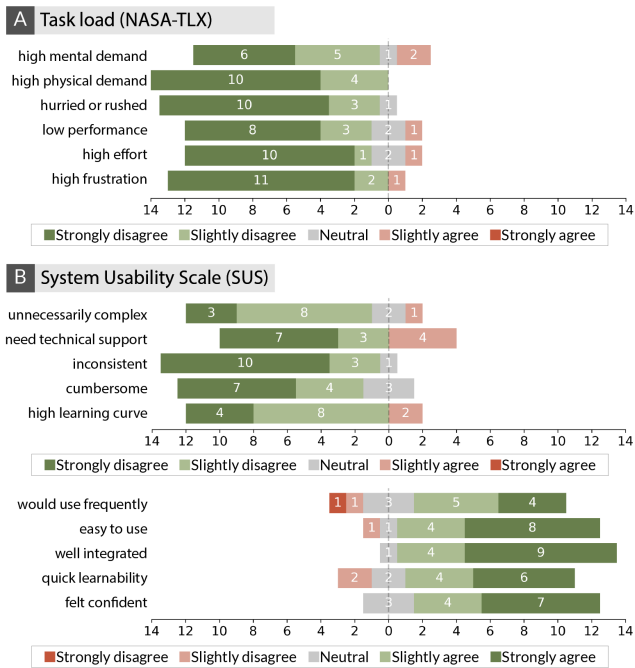

Figure 11: (A) NASA Task Load Index and (B) System Usability.

**Users' mental model of abstraction.** Our system abstracts robot capabilities into object-oriented actions, which in our study appeared to shape users' mental models. Participants referred to "the robot" when describing tasks without distinguishing between individual robots' capabilities. While emphasizing object-level affordances can facilitate fluid interactions, it may obscure differences between robots and limit users' ability to delegate tasks or anticipate failures. This trade-off warrants further study as robots become more specialized. Action abstraction is also a key design challenge. Low-level capabilities (e.g., navigation, joint movement) are encapsulated into grasping, then wrapped into pick-and-place and presented to the user as a *Move* action. Higher-level composite actions such as *Fetch*, *Stack*, or *Sort* embed multiple pick-and-place substeps. Rather than exposing low-level commands, X-OOHRI presents user-centered action possibilities, with carefully chosen abstraction levels and support for user-defined composite actions [57]. Further research is needed to understand users' mental model of abstraction.

**Automation and user control.** A core design principle is to prioritize user control. Even when the robot could autonomously resolve a constraint, users have to select the Auto option, preview, and confirm before execution. Ideally, this user oversight should only be necessary when the robot is uncertain about its autonomous actions [68]. The design of X-OOHRI could be further explored to support higher levels of automation alongside user control [74]. While our focus on communication emphasized explicit commands, X-OOHRI could also make autonomous actions more transparent during implicit interactions and proactive behaviors [46, 64, 84].

## 6 Limitations and Future Work

We highlight some shortcomings of X-OOHRI and discuss opportunities for extending it along several dimensions.

**VLM errors.** We identified visual errors from misestimated attributes for non-observable or deceptive properties (e.g., the weight of an opaque bottle or a fake book) and spatial errors caused by occlusion, clutter, or limited field of view. While visual errors are inherent to VLMs, spatial errors can be reduced in AR through multi-view capture and closed-loop execution feedback that updates object attributes, enabling a self-correcting process.

**Remote teleoperation.** Our system instantiates X-OOHRI in AR, emphasizing spatial alignment with physical counterparts and assuming user colocation. Yet X-OOHRI is not limited to AR; with high-fidelity 3D reconstructions, it could be extended to VR, allowing OO interactions with photorealistic scenes across multiple locations (e.g., package delivery). Future advances in 4D Gaussian Splatting may further enable dynamic VR interactions from the robot's perspective, supporting remote real-time teleoperation.

**Further evaluation.** Our preliminary evaluation can be extended with more complex task sequences and highly abstract instructions for advanced robots. Future work should also examine X-OOHRI in longitudinal in-home studies and investigate how object-oriented communication shapes user trust and reliance as robot capabilities evolve or failures occur.

**Robot learning.** X-OOHRI primarily serves as an AR interface for user commands and explanations, but it could also inform robot learning. Constraining direct manipulation to feasible trajectories may reveal user preferences, though at the cost of fluidity, while object-oriented interactions and final scene states could be leveraged to learn and evaluate task success [43, 55].

## 7 Conclusion

In this paper, we present X-OOHRI, a novel AR paradigm that conveys robot capabilities and limitations during user instruction through a set of communication channels, including signifiers, action and resolution menus, color codings, explanation tags, and preview animations. Our system encodes object properties and robot limits as object-oriented affordances via a VLM and integrates them with environment bounds to construct a spatially aligned simulated environment. It enables direct manipulation for personalized instructions, real-time explanation updates, and end-to-end robot execution. Through technical demonstrations across example scenarios, we highlight the practical usability and envision future applications across diverse domains and integration with physical robots. Finally, results from our study show that users effectively issued commands through object-oriented interactions, reporting high levels of integration and ease of use. Participants demonstrated a clear understanding of the robots' abilities and constraints and identified situations requiring user support. Together, these findings underscore X-OOHRI's potential and open new possibilities for human–robot interaction.

## Acknowledgments
We thank Karthik Mahadevan for his feedback. This research was supported by the Princeton NextG Innovation Grant and the Princeton SEAS Innovation Fund.

# References

[1] Michael Ahn, Anthony Brohan, Noah Brown, Yevgen Chebotar, Omar Cortes, Byron David, Chelsea Finn, Chuyuan Fu, Keerthana Gopalakrishnan, Karol Hausman, Alex Herzog, Daniel Ho, Jasmine Hsu, Julian Ibarz, Brian Ichter, Alex Irpan, Eric Jang, Rosario Jauregui Ruano, Kyle Jeffrey, Sally Jesmonth, Nikhil Joshi, Ryan Julian, Dmitry Kalashnikov, Yuheng Kuang, Kuang-Huei Lee, Sergey Levine, Yao Lu, Linda Luu, Carolina Parada, Peter Pastor, Jornell Quiambao, Kanishka Rao, Jarek Rettinghouse, Diego Reyes, Pierre Sermanet, Nicolas Sievers, Clayton Tan, Alexander Toshev, Vincent Vanhoucke, Fei Xia, Ted Xiao, Peng Xu, Sichun Xu, Mengyuan Yan, and Andy Zeng. 2022. Do As I Can and Not As I Say: Grounding Language in Robotic Affordances. In *Proceedings of the 6th Conference on Robot Learning*. PMLR, 287–318. https://proceedings.mlr.press/v205/ichter23a.html

[2] Gopika Ajaykumar, Maureen Steele, and Chien-Ming Huang. 2021. A Survey on End-User Robot Programming. *Comput. Surveys* 54, 8 (2021). doi:10.1145/3466819

[3] Anesh Alvanpour, Sumit Kumar Das, Christopher Kevin Robinson, Olfa Nasraoui, and Dan Popa. 2020. Robot Failure Mode Prediction with Explainable Machine Learning. In *Proceedings of the 2020 IEEE 16th International Conference on Automation Science and Engineering*. IEEE, 61–66. doi:10.1109/CASE48305.2020.9216965

[4] Saleema Amershi, Ece Kamar, and Emre Kiciman. 2019. People and AI See Things Different: Implications of Mismatched Perception on HCI for AI Systems. In *Workshop on Human-Centered Machine Learning Perspectives at the 2019 CHI Conference on Human Factors in Computing Systems*. 1–6. https://www.microsoft.com/en-us/research/publication/people-and-ai-see-things-different-implications-of-mismatched-perception-on-hci-for-ai-systems/

[5] Rasmus S. Andersen, Ole Madsen, Thomas B. Moeslund, and Heni Ben Amor. 2016. Projecting Robot Intentions into Human Environments. In *Proceedings of the 2016 25th IEEE International Symposium on Robot and Human Interactive Communication*. IEEE, 294–301. doi:10.1109/ROMAN.2016.7745145

[6] Shigeto Aramaki. 1987. Object Oriented Approach for Task Level Robot Programming. In *Proceedings of the 1987 IEEE International Conference on Industrial Electronics, Control, and Instrumentation: Industrial Applications of Robotics Machine Vision*. SPIE, 596–601. doi:10.1117/12.943014

[7] Shigeto Aramaki. 1991. Knowledge Based Robot Programming. *Journal of Robotics and Mechatronics* 3, 5 (1991), 428–434. doi:10.20965/jrm.1991.p0428

[8] Stephanie Arevalo Arboleda, Franziska Rücker, Tim Dierks, and Jens Gerken. 2021. Assisting Manipulation and Grasping in Robot Teleoperation with Augmented Reality Visual Cues. In *Proceedings of the 2021 CHI Conference on Human Factors in Computing Systems*. ACM, 1–14. doi:10.1145/3411764.3445398

[9] Deborah J. Armstrong. 2006. The Quarks of Object-Oriented Development. *Commun. ACM* 49, 2 (2006), 123–128. doi:10.1145/1113034.1113040

[10] Watanabe Atsushi, Tetsushi Ikeda, Yoichi Morales, Kazuhiko Shinozawa, Takahiro Miyashita, and Norihiro Hagita. 2015. Communicating Robotic Navigational Intentions. In *Proceedings of the 2015 IEEE/RSJ International Conference on Intelligent Robots and Systems*. IEEE, 5763–5769. doi:10.1109/IROS.2015.7354195

[11] Andrea Bauer, Dirk Wollherr, and Martin Buss. 2008. Human-Robot Collaboration: A Survey. *International Journal of Humanoid Robotics* 5 (2008), 47–66. doi:10.1142/S0219843608001303

[12] Arne Bilberg and Ali Ahmad Malik. 2019. Digital Twin Driven Human-Robot Collaborative Assembly. *CIRP Annals* 68, 1 (2019), 499–502. doi:10.1016/j.cirp.2019.04.011

[13] Darcy M. Bullock and Irving J. Oppenheim. 1992. Object-Oriented Programming in Robotics Research for Excavation. *Journal of Computing in Civil Engineering* 6, 3 (1992), 370–385. doi:10.1061/(ASCE)0887-3801(1992)6:3(370)

[14] Xuan Cao, Jacob W Crandall, and Michael A Goodrich. 2024. Generating Explanations for Autonomous Robots Using Assumption-Alignment Tracking. In *Proceedings of the 2024 IEEE International Conference on Systems, Man, and Cybernetics*. IEEE, 1169–1174. doi:10.1109/SMC54092.2024.10831560

[15] Yuanzhi Cao, Tianyi Wang, Xun Qian, Pawan S. Rao, Manav Wadhawan, Ke Huo, and Karthik Ramani. 2019. GhostAR: A Time-space Editor for Embodied Authoring of Human-Robot Collaborative Task with Augmented Reality. In *Proceedings of the 32nd Annual ACM Symposium on User Interface Software and Technology*. ACM, 521–534. doi:10.1145/3332165.3347902

[16] Yuanzhi Cao, Zhuangying Xu, Fan Li, Wentao Zhong, Ke Huo, and Karthik Ramani. 2019. V.Ra: An In-Situ Visual Authoring System for Robot-IoT Task Planning with Augmented Reality. In *Proceedings of the 2019 Conference on Designing Interactive Systems*. ACM, 1059–1070. doi:10.1145/3322276.3322278

[17] Rodrigo Chacon-Quesada and Yiannis Demiris. 2020. Augmented Reality User Interfaces for Heterogeneous Multirobot Control. In *Proceedings of the 2020 IEEE/RSJ International Conference on Intelligent Robots and Systems (IROS)*. IEEE, 11439–11444. doi:10.1109/IROS45743.2020.9341422

[18] Nicholas Conlon, Aastha Acharya, Jamison McGinley, Trevor Slack, Camron A Hirst, Marissa D'Alonzo, Mitchell R Hebert, Christopher Reale, Eric W Frew, Rebecca Russell, et al. 2022. Generalizing Competency Self-Assessment for Autonomous Vehicles Using Deep Reinforcement Learning. In *Proceedings of the 2022 AIAA SciTech Forum*. AIAA, 1–20. doi:10.2514/6.2022-2496

[19] Nicholas Conlon, Nisar R Ahmed, and Daniel Szafir. 2024. A Survey of Algorithmic Methods for Competency Self-Assessments in Human-Autonomy Teaming. *Comput. Surveys* 56, 7 (2024), 1–31. doi:10.1145/3616010

[20] Nicholas Conlon, Daniel Szafir, and Nisar Ahmed. 2022. "I'm Confident This Will End Poorly": Robot Proficiency Self-Assessment in Human-Robot Teaming. In *Proceedings of the 2022 IEEE/RSJ International Conference on Intelligent Robots and Systems*. IEEE, 2127–2134. doi:10.1109/IROS47612.2022.9981653

[21] Eric Corbett and Astrid Weber. 2016. What Can I Say? Addressing User Experience Challenges of a Mobile Voice User Interface for Accessibility. In *Proceedings of the 18th International Conference on Human-Computer Interaction with Mobile Devices and Services*. ACM, 72–82. doi:10.1145/2935334.2935386

[22] Erol Şahin, Maya Çakmak, Mehmet R. Doğar, Emre Uğur, and Göktürk Üçoluk. 2007. To Afford or Not to Afford: A New Formalization of Affordances Toward Affordance-Based Robot Control. *Adaptive Behavior* 15, 4 (2007), 447–472. doi:10.1177/1059712307084689

[23] Devleena Das, Siddhartha Banerjee, and Sonia Chernova. 2021. Explainable AI for Robot Failures: Generating Explanations that Improve User Assistance in Fault Recovery. In *Proceedings of the 2021 ACM/IEEE International Conference on Human-Robot Interaction*. ACM, 351–360. doi:10.1145/3434073.3444657

[24] Fatih Dogangun, Serdar Bahar, Yigit Yildirim, Bora Toprak Temir, Emre Ugur, and Mustafa Doga Dogan. 2025. RAMPA: Robotic Augmented Reality for Machine Programming by DemonstrAtion. *IEEE Robotics and Automation Letters* 10, 4 (2025), 3795–3802. doi:10.1109/lra.2025.3546109

[25] Mauro Dragone, Brian R Duffy, Thomas Holz, and Greg MP O'Hare. 2006. Fusing Realities in Human-Robot Social Interaction. In *Proceedings of the 37th International Symposium on Robotics*. VDI, 1–14.

[26] Jiafei Duan, Wilbert Pumacay, Nishanth Kumar, Yi Ru Wang, Shulin Tian, Wentao Yuan, Ranjay Krishna, Dieter Fox, Ajay Mandlekar, and Yijie Guo. 2026. AHA: A Vision-Language-Model for Detecting and Reasoning over Failures in Robotic Manipulation. In *Proceedings of the 14th International Conference on Learning Representations*. 1–16.

[27] H. C. Fang, S. K. Ong, and A. Y. C. Nee. 2012. Interactive Robot Trajectory Planning and Simulation Using Augmented Reality. *Robotics and Computer-Integrated Manufacturing* 28, 2 (2012), 227–237. doi:10.1016/j.rcim.2011.09.003

[28] Samir Yitzhak Gadre, Eric Rosen, Gary Chien, Elizabeth Phillips, Stefanie Tellex, and George Konidaris. 2019. End-User Robot Programming Using Mixed Reality. In *Proceedings of the 2019 International Conference on Robotics and Automation*. IEEE, 2707–2713. doi:10.1109/ICRA.2019.8793988

[29] Pablo Soler Garcia, Petar Lukovic, Lucie Reynaud, Andrea Sgobbi, Federica Bruni, Martin Brun, Marc Zünd, Riccardo Bollati, Marc Pollefeys, Hermann Blum, and Zuria Bauer. 2024. HoloSpot: Intuitive Object Manipulation via Mixed Reality Drag-and-Drop. *arXiv preprint arXiv:2410.11110* (2024). https://arXiv.org/abs/2410.11110

[30] Alvika Gautam, Tim Whiting, Xuan Cao, Michael A Goodrich, and Jacob W Crandall. 2022. A Method for Designing Autonomous Robots That Know Their Limits. In *Proceedings of the 2022 International Conference on Robotics and Automation*. IEEE, 121–127. doi:10.1109/ICRA46639.2022.9812030

[31] McKee G.T., J.A. Fryer, and P.S. Schenker. 2001. Object-Oriented Concepts for Modular Robotics Systems. In *Proceedings of the 39th IEEE International Conference and Exhibition on Technology of Object-Oriented Languages and Systems*. IEEE, 229–238. doi:10.1109/TOOLS.2001.941676

[32] Wortham Robert H, Andreas Theodorou, and Bryson Joanna J. 2016. What Does the Robot Think? Transparency as a Fundamental Design Requirement for Intelligent Systems. In *Workshop on Ethics for Artificial Intelligence at the 2016 International Joint Conference on Artificial Intelligence*. AAAI, 1–6.

[33] Andrei Haidu, Daniel Kohlsdorf, and Michael Beetz. 2015. Learning Action Failure Models from Interactive Physics-Based Simulations. In *Proceedings of the 2015 IEEE/RSJ International Conference on Intelligent Robots and Systems*. IEEE, 5370–5375. doi:10.1109/IROS.2015.7354136

[34] Zhao Han, Yifei Zhu, Albert Phan, Fernando Sandoval Garza, Amia Castro, and Tom Williams. 2023. Crossing Reality: Comparing Physical and Virtual Robot Deixis. In *Proceedings of the 2023 ACM/IEEE International Conference on Human-Robot Interaction*. ACM, 152–161. doi:10.1145/3568162.3576972

[35] Ziyao He, Yunpeng Song, Shurui Zhou, and Zhongmin Cai. 2023. Interaction of Thoughts: Towards Mediating Task Assignment in Human-AI Cooperation with a Capability-Aware Shared Mental Model. In *Proceedings of the 2023 CHI Conference on Human Factors in Computing Systems*. ACM, 1–18. doi:10.1145/3544548.3580983

[36] Hooman Hedayati, Michael Walker, and Daniel Szafir. 2018. Improving Collocated Robot Teleoperation with Augmented Reality. In *Proceedings of the 2018 ACM/IEEE International Conference on Human-Robot Interaction*. ACM, 78–86. doi:10.1145/3171221.3171251

[37] Valentin Heun, James Hobin, and Pattie Maes. 2013. Reality Editor: Programming Smarter Objects. In *Proceedings of the 2013 ACM Conference on Pervasive and Ubiquitous Computing Adjunct Publication*. ACM, 307–310. doi:10.1145/2494091.2494185

[38] Valentin Heun, Shunichi Kasahara, and Pattie Maes. 2013. Smarter Objects: Using AR Technology to Program Physical Objects and Their Interactions. In *Extended Abstracts of the 2013 CHI Conference on Human Factors in Computing Systems*.

ACM, 961–966. doi:10.1145/2468356.2468528
[39] Shanee Honig and Tal Oron-Gilad. 2018. Understanding and Resolving Failures in Human-Robot Interaction: Literature Review and Model Development. *Frontiers in Psychology* 9 (2018). doi:10.3389/fpsyg.2018.00861
[40] Bryce Ikeda, Maitrey Gramopadhye, LillyAnn Nekervis, and Daniel Szafir. 2025. MARCER: Multimodal Augmented Reality for Composing and Executing Robot Tasks. In *Proceedings of the 2025 ACM/IEEE International Conference on Human-Robot Interaction*. ACM/IEEE, 529–539. doi:10.1109/HRI61500.2025.10974232
[41] Brett Israelsen, Nisar R Ahmed, Matthew Aitken, Eric W Frew, Dale A Lawrence, and Brian M Argrow. 2024. "A Good Bot Always Knows Its Limitations": Assessing Autonomous System Decision-making Competencies through Factorized Machine Self-confidence. *ACM Transactions on Human-Robot Interaction* 14, 4 (2024). doi:10.1145/3732794
[42] Hyun Joe Jeong and Andrea Bajcsy. 2024. Robots That Suggest Safe Alternatives. *arXiv preprint arXiv:2409.09883* (2024). https://arXiv.org/abs/2409.09883
[43] Yunfan Jiang, Agrim Gupta, Zichen Zhang, Guanzhi Wang, Yongqiang Dou, Yanjun Chen, Li Fei-Fei, Anima Anandkumar, Yuke Zhu, and Linxi Fan. 2023. VIMA: General Robot Manipulation with Multimodal Prompts. In *Proceedings of the 40th International Conference on Machine Learning*. PMLR. https://proceedings.mlr.press/v202/jiang23b.html
[44] A.K. Jones and B.H. Liskov. 1976. A Language Extension for Controlling Access to Shared Data. *IEEE Transactions on Software Engineering* SE-2, 4 (1976), 277–285. doi:10.1109/TSE.1976.233833
[45] Modayil Joseph and Kuipers Benjamin. 2008. The Initial Development of Object Knowledge by a Learning Robot. *Robotics and Autonomous Systems* 56, 11 (2008), 879–890. doi:10.1016/j.robot.2008.08.004
[46] Wendy Ju. 2015. *The Design of Implicit Interactions*. Springer Nature. https://www.worldcat.org/isbn/9783031022104
[47] Mohamed Kari and Parastoo Abtahi. 2025. Reality Promises: Virtual-Physical Decoupling Illusions in Mixed Reality via Invisible Mobile Robots. In *Proceedings of the 38th Annual ACM Symposium on User Interface Software and Technology*. ACM, 1–17. doi:10.1145/3746059.3747660
[48] Alan Kay. 2003. Dr. Alan Kay on the Meaning of "Object-Oriented Programming". Email to Stefan L. Ram, archived on FU Berlin website. https://www.purl.org/stefan_ram/pub/doc_kay_oop_en Accessed via Stefan L. Ram's webpage.
[49] S. Kiesler. 2005. Fostering Common Ground in Human-Robot Interaction. In *Workshop at the 2005 IEEE Robot and Human Interactive Communication*. IEEE, 729–734. doi:10.1109/ROMAN.2005.1513866
[50] Sunnie S. Y. Kim, Elizabeth Anne Watkins, Olga Russakovsky, Ruth Fong, and Andrés Monroy-Hernández. 2023. "Help Me Help the AI": Understanding How Explainability Can Support Human-AI Interaction. In *Proceedings of the 2023 CHI Conference on Human Factors in Computing Systems*. ACM, 1–17. doi:10.1145/3544548.3581001
[51] Benjamin Lee, Michael Sedlmair, and Dieter Schmalstieg. 2023. Design Patterns for Situated Visualization in Augmented Reality. *IEEE Transactions on Visualization and Computer Graphics* 30, 1 (2023), 1–12. doi:10.1109/TVCG.2023.3327398
[52] Sau-lai Lee, Ivy Yee-man Lau, S. Kiesler, and Chi-Yue Chiu. 2005. Human Mental Models of Humanoid Robots. In *Proceedings of the 2005 IEEE International Conference on Robotics and Automation*. IEEE, 2767–2772. doi:10.1109/ROBOT.2005.1570532
[53] Sang-won Leigh and Pattie Maes. 2015. AfterMath: Visualizing Consequences of Actions through Augmented Reality. In *Extended Abstracts of the 2015 CHI Conference on Human Factors in Computing Systems*. ACM, 941–946. doi:10.1145/2702613.2732695
[54] Rasmus Skovhus Lunding, Mille Skovhus Lunding, Tiare Feuchtner, Marianne Graves Petersen, Kaj Grønbæk, and Ryo Suzuki. 2024. RoboVisAR: Immersive Authoring of Condition-based AR Robot Visualisations. In *Proceedings of the 2024 ACM/IEEE International Conference on Human-Robot Interaction*. ACM, 462–471. doi:10.1145/3610977.3634972
[55] Chenyang Ma, Yue Yang, Bryce Ikeda, and Daniel Szafir. 2025. Supporting Long-Horizon Tasks in Human-Robot Collaboration by Aligning Intentions via Augmented Reality. In *Proceedings of the 2025 ACM/IEEE International Conference on Human-Robot Interaction*. IEEE, 1468–1472. doi:10.1109/HRI61500.2025.10973947
[56] Oscar Mañas, Benno Krojer, and Aishwarya Agrawal. 2024. Improving Automatic VQA Evaluation Using Large Language Models. In *Proceedings of the 38th Conference on Artificial Intelligence and 36th Conference on Innovative Applications of Artificial Intelligence and 14th Symposium on Educational Advances in Artificial Intelligence*. AAAI, 1–9. doi:10.1609/aaai.v38i5.28212
[57] Karthik Mahadevan, Yan Chen, Maya Cakmak, Anthony Tang, and Tovi Grossman. 2022. Mimic: In-Situ Recording and Re-Use of Demonstrations to Support Robot Teleoperation. In *Proceedings of the 35th Annual ACM Symposium on User Interface Software and Technology*. ACM, 1–13. doi:10.1145/3526113.3545639
[58] Bernardo Marques, Beatriz Sousa Santos, Tiago Araújo, Nuno Cid Martins, João Bernardo Alves, and Paulo Dias. 2019. Situated Visualization in the Decision Process Through Augmented Reality. In *Proceedings of the 2019 23rd International Conference Information Visualisation*. IEEE, 13–18. doi:10.1109/IV.2019.00012
[59] Andreea Muresan, Jess Mcintosh, and Kasper Hornbæk. 2023. Using Feedforward to Reveal Interaction Possibilities in Virtual Reality. *ACM Transactions on Computer-Human Interaction* 30, 6 (2023). doi:10.1145/3603623
[60] Kawa Nazemi. 2018. *Adaptive Semantics Visualization*. Springer. https://www.worldcat.org/isbn/9783319808932
[61] D.A. Norman. 1988. *The Psychology of Everyday Things*. Basic Books. https://www.worldcat.org/isbn/9780385267748
[62] Scott Ososky, David Schuster, Florian Jentsch, Stephen Fiore, Randall Shumaker, Christian Lebiere, Unmesh Kurup, Jean Oh, and Anthony Stentz. 2012. The Importance of Shared Mental Models and Shared Situation Awareness for Transforming Robots from Tools to Teammates. In *Proceedings of the 2012 SPIE Conference on Defense, Security, and Sensing: Unmanned Systems Technology XIV*. SPIE, 1–12. doi:10.1117/12.923283
[63] Max Pascher, Uwe Gruenefeld, Stefan Schneegass, and Jens Gerken. 2023. How to Communicate Robot Motion Intent: A Scoping Review. In *Proceedings of the 2023 CHI Conference on Human Factors in Computing Systems*. ACM, 1–17. doi:10.1145/3544548.3580857
[64] Maithili Patel and Sonia Chernova. 2022. Proactive Robot Assistance via Spatio-Temporal Object Modeling. In *Proceedings of the 6th Conference on Robot Learning*. PMLR, 881–891. https://proceedings.mlr.press/v205/patel23a.html
[65] Rodrigo Chacón Quesada and Yiannis Demiris. 2022. Proactive Robot Assistance: Affordance-Aware Augmented Reality User Interfaces. *IEEE Robotics & Automation Magazine* 29, 1 (2022), 22–34. doi:10.1109/MRA.2021.3136789
[66] Camilo Perez Quintero, Sarah Li, Matthew KXJ Pan, Wesley P. Chan, H.F. Machiel Van der Loos, and Elizabeth Croft. 2018. Robot Programming Through Augmented Trajectories in Augmented Reality. In *Proceedings of the 2018 IEEE/RSJ International Conference on Intelligent Robots and Systems*. IEEE, 1838–1844. doi:10.1109/IROS.2018.8593700
[67] Gourdeau R. 1997. Object-Oriented Programming for Robotic Manipulator Simulation. *IEEE Robotics & Automation Magazine* 4, 3 (1997), 21–29. doi:10.1109/100.618020
[68] Allen Z. Ren, Anushri Dixit, Alexandra Bodrova, Sumeet Singh, Stephen Tu, Noah Brown, Peng Xu, Leila Takayama, Fei Xia, Jake Varley, Zhenjia Xu, Dorsa Sadigh, Andy Zeng, and Anirudha Majumdar. 2023. Robots That Ask For Help: Uncertainty Alignment for Large Language Model Planners. In *Proceedings of the 7th Conference on Robot Learning*. PMLR, 661–682. https://proceedings.mlr.press/v229/ren23a.html
[69] Eric Rosen, David Whitney, Elizabeth Phillips, Gary Chien, James Tompkin, George Konidaris, and Stefanie Tellex. 2020. Communicating Robot Arm Motion Intent Through Mixed Reality Head-Mounted Displays. In *Proceedings of the 2020 Springer International Symposium on Robotics Research*. Springer, 301–316. doi:10.1007/978-3-030-28619-4_26
[70] Stephanie Rosenthal, Sai P Selvaraj, and Manuela M Veloso. 2016. Verbalization: Narration of Autonomous Robot Experience. In *Proceedings of the 2016 International Joint Conference on Artificial Intelligence*. 862–868.
[71] Som Sagar, Jiafei Duan, Sreevishakh Vasudevan, Yifan Zhou, Heni Ben Amor, Dieter Fox, and Ransalu Senanayake. 2025. From Mystery to Mastery: Failure Diagnosis for Improving Manipulation Policies. In *Workshop on Out-of-Distribution Generalization in Robotics at the 2025 Conference on Robotics: Science and Systems*. 1–17.
[72] Yasaman S. Sefidgar, Thomas Weng, Heather Harvey, Sarah Elliott, and Maya Cakmak. 2018. RobotIST: Interactive Situated Tangible Robot Programming. In *Proceedings of the 2018 ACM Symposium on Spatial User Interaction*. ACM, 141–149. doi:10.1145/3267782.3267921
[73] Rossitza Setchi, Maryam Banitalebi Dehkordi, and Juwairiya Siraj Khan. 2020. Explainable Robotics in Human-Robot Interactions. *Procedia Computer Science* 176 (2020), 3057–3066. doi:10.1016/j.procs.2020.09.198
[74] Ben Shneiderman. 2020. Human-Centered Artificial Intelligence: Reliable, Safe and Trustworthy. *International Journal of Human-Computer Interaction* 36, 6 (2020), 495–504. doi:10.1080/10447318.2020.1741118
[75] Maia Stiber and Chien-Ming Huang. 2021. Not All Errors Are Created Equal: Exploring Human Responses to Robot Errors with Varying Severity. In *Late Breaking Results of the 2020 ACM International Conference on Multimodal Interaction*. ACM, 97–101. doi:10.1145/3395035.3425245
[76] Ryo Suzuki, Adnan Karim, Tian Xia, Hooman Hedayati, and Nicolai Marquardt. 2022. Augmented Reality and Robotics: A Survey and Taxonomy for AR-enhanced Human-Robot Interaction and Robotic Interfaces. In *Proceedings of the 2022 CHI Conference on Human Factors in Computing Systems*. ACM, 1–3. doi:10.1145/3491102.3517719
[77] Michael Walker, Hooman Hedayati, Jennifer Lee, and Daniel Szafir. 2018. Communicating Robot Motion Intent with Augmented Reality. In *Proceedings of the 2018 ACM/IEEE International Conference on Human-Robot Interaction*. ACM, 316–324. doi:10.1145/3171221.3171253
[78] Chao Wang, Anna Belardinelli, Stephan Hasler, Theodoros Stouraitis, Daniel Tanneberg, and Michael Gienger. 2023. Explainable Human-Robot Training and Cooperation with Augmented Reality. In *Extended Abstracts of the 2023 CHI Conference on Human Factors in Computing Systems*. ACM, 1–5. doi:10.1145/3544549.3583889
[79] Lauren W. Wang and Parastoo Abtahi. 2025. GhostObjects: Instructing Robots by Manipulating Spatially Aligned Virtual Twins in Augmented Reality. In *Adjunct*

*Proceedings of the 38th Annual ACM Symposium on User Interface Software and Technology*. ACM. doi:10.1145/3746058.3758451

[80] Xi Wang, Ci-Jyun Liang, Carol C Menassa, and Vineet R Kamat. 2021. Interactive and Immersive Process-Level Digital Twin for Collaborative Human–Robot Construction Work. *Journal of Computing in Civil Engineering* 35, 6 (2021), 04021023. doi:10.1061/(ASCE)CP.1943-5487.0000988

[81] Zihan Wang, Brian Liang, Varad Dhat, Zander Brumbaugh, Nick Walker, Ranjay Krishna, and Maya Cakmak. 2024. I Can Tell What I am Doing: Toward Real-World Natural Language Grounding of Robot Experiences. In *Proceedings of the 8th Conference on Robot Learning*. PMLR, 1863–1890. https://proceedings.mlr.press/v270/wang25g.html

[82] Wesley Willett, Yvonne Jansen, and Pierre Dragicevic. 2017. Embedded Data Representations. *IEEE Transactions on Visualization and Computer Graphics* 23, 1 (2017), 461–470. doi:10.1109/TVCG.2016.2598608

[83] Michal Yarom, Yonatan Bitton, Soravit Changpinyo, Roee Aharoni, Jonathan Herzig, Oran Lang, Eran Ofek, and Idan Szpektor. 2023. What You See is What You Read? Improving Text-Image Alignment Evaluation. In *Advances in Neural Information Processing Systems*. Curran, 1601–1619. https://papers.nips.cc/paper_files/paper/2023/hash/056e8e9c8ca9929cb6cf198952bf1dbb-Abstract-Conference.html

[84] Wentao Yuan, Jiafei Duan, Valts Blukis, Wilbert Pumacay, Ranjay Krishna, Adithyavairavan Murali, Arsalan Mousavian, and Dieter Fox. 2024. RoboPoint: A Vision-Language Model for Spatial Affordance Prediction for Robotics. In *Proceedings of the 8th Conference on Robot Learning*. PMLR, 4005–4020. https://proceedings.mlr.press/v270/yuan25c.html

[85] Georgios A. Zachiotis, George Andrikopoulos, Randy Gornez, Keisuke Nakamura, and George Nikolakopoulos. 2018. A Survey on the Application Trends of Home Service Robotics. In *Proceedings of the 2018 IEEE International Conference on Robotics and Biomimetics*. IEEE, 1999–2006. doi:10.1109/ROBIO.2018.8665127

[86] Danny Zhu. 2017. *Augmented Reality Visualization for Autonomous Robots*. Ph. D. Dissertation. Air Force Research Laboratory. http://reports-archive.adm.cs.cmu.edu/anon/2017/CMU-CS-17-129.pdf